\documentclass[copyright]{eptcs}
\providecommand{\event}{T. Neary, D. Woods, A.K. Seda and N. Murphy (Eds.):\\ The Complexity of Simple Programs 2008.}
\providecommand{\volume}{1}
\providecommand{\anno}{2009}
\providecommand{\firstpage}{56}
\usepackage{breakurl}
\usepackage{graphicx,amssymb}
\usepackage[english]{babel}
\usepackage{amsfonts}
\usepackage{longtable}

\newtheorem{normalsystem}{Definition}
\newtheorem{tagsystem}[normalsystem]{Definition}
\newtheorem{halt}[normalsystem]{Definition}
\newtheorem{reach}[normalsystem]{Definition}

\title{On the boundaries of solvability and unsolvability in tag systems. Theoretical and Experimental Results.}
\author{Liesbeth De Mol
    \institute{Gent University, Centre for Logic and Philosophy of Science, Blandijnberg 2, 9000 Gent, Belgium.\thanks{This research was supported by the Flemish fund for Scientific Research --  FWO, Belgium and the Kunsthochschule f\"ur Medien, K\"oln.}}
    \email{elizabeth.demol@ugent.be}
}
\def\titlerunning{On the boundaries of solvability and unsolvability in tag systems}
\def\authorrunning{L. De Mol}

\begin{document}
\maketitle

\begin{abstract}
Several older and more recent results on the boundaries of solvability and unsolvability in tag systems are surveyed. Emphasis will be put on the significance of computer experiments in research on very small tag systems. 
\end{abstract}
\section{Introduction}\label{Intro}
Tag systems were invented and studied by Emil Leon Post during his Procter fellowship at Princeton during the academic year 1920-21 \cite{post43, post65}. Although Post never wanted to work on these systems again, he was convinced that they are recursively unsolvable. This was proven by Minsky in 1961 \cite{minsk61a}. He showed that any Turing machine can be reduced to a tag system with $v = 6$, where  $v$  denotes the number of deleted letters (see Definition 1, p. \pageref{DefTag}). This result was improved by Cocke and Minsky to $v = 2$ \cite{CockeMinsky64,minsk62}.\\ 
Minsky was the first to construct a very small universal Turing machine that simulates 2-tag systems \cite{minsk62}. The machine has 7 states and 4 symbols. Minsky's machine, however, has a defect: it erases its output before halting. Hence, it is not a truly universal machine. Nowadays, tag systems still play a fundamental role in the race for finding small universal systems. I.e., many small universal systems are proven to be universal through simulation of 2-tag systems, or some variant of tag systems. For example, Rogozhin \cite{rog82} constructed several small universal Turing machines by 2-tag simulation and improved Minsky's machine. Neary and Woods \cite{near06} found a universal Turing machine with 6 states and 4 symbols, simulating what they have called bi-tag systems, a variant of tag systems. For an overview on small universal systems we refer the reader to \cite{marg00}.\\
Given, on the one hand, the simplicity of the form of tag systems, and, on the other hand, their computational power, tag systems are also possible candidates for finding the ``simplest'' possible universal system. Because of this simplicity, they are also very suitable to be implemented on and studied with the help of the computer. These reasons motivate more research on tag systems for their own sake.\\
In this paper we will survey and discuss several older and more recent results on tag systems.  The main focus of the paper are results connected to finding small universal tag systems. We will consider two main approaches: a more theoretical (Sec.~\ref{TheorResults}) and a more heuristic approach (Sec.~\ref{HeuristicResults}). Emphasis will be put on the significance of experimental research on very small tag systems, the largest portion of this paper being devoted to  
a summary of previously unpublished results from computer experiments on tag systems. The motivation behind these experiments is that, for now, complementing theoretical with more experimental methods seems to be the best method available to come to a better understanding of the complex behavior of very small tag systems.

\subsection{Definitions and notational conventions}\label{definitions}
 \begin{tagsystem}\label{DefTag}
A tag system $T$ consists of a finite alphabet $\Sigma$ of $\mu$ symbols, a number $v \in \mathbb{N}$ and a finite set of $\mu$ words $w_{0}, w_{1},\ldots,w_{\mu-1} \in \Sigma^{\ast}$, called the appendants, where any word $w_{i}$ corresponds with $a_{i} \in \Sigma$.
\end{tagsystem}
Given an initial word $A_0 \in \Sigma^{\ast}$, the tag system appends the appendant $w_{i}$
associated with the leftmost letter $a_{0, i}$ of $A_0$ at the end of $A_0$,
and deletes the first $v$ symbols of $A_0$. This computational process is iterated until the tag system halts, i.e. produces the empty word $\epsilon$. If this
does not happen the tag system can either become periodic or show
unbounded growth.\\ 
To give an example, let us consider the tag system mentioned by Post with $v = 3$, $0 \rightarrow 00$, $1 \rightarrow 1101$ \cite{post43}. If $A_{0} = 001101$ we get:\\
\hspace*{0.7cm}\textbf{001}101\\
\hspace*{1.4cm}\textbf{101}00\\
\hspace*{2.1cm}\textbf{001}101\\
The word $A_{0}$ is reproduced after 2 computation steps and is thus an example of a periodic word. Post called the behaviour of this tag system ``intractable''. He also mentioned that he studied the class of tag systems with $v = 2, \mu = 3$ and described it as being of ``\textit{bewildering complexity}'' and as ``\textit{[\ldots] leading to an overwhelming confusion of classes of cases, with the solution of the corresponding problem depending more and more on problems of ordinary number theory.}'' \cite{post65}.\\
Post considered two decision problems for tag systems, which we will call the
\textit{halting problem} and the \textit{reachability problem} for tag systems.
\begin{halt}
The halting problem for tag systems is the problem to determine
for a given tag system $T$ and initial word $A_{0}$ whether or not $T$ will halt when started from $A_{0}$.
\end{halt}
\begin{reach}The reachability problem for tag systems is the problem to determine for a
given tag system $T$, a given initial word $A_{0}$ and an
arbitrary word $A \inÊ\Sigma^{\ast}$, whether or not $T$ will
ever produce $A$ when started from $A_{0}$.
\end{reach}
Note that the halting problem is a special case of the reachability problem.\\
In the remainder of the paper we will use the notations and definitions given in this paragraph. 
In what follows, TS$(\mu, v)$ denotes the class of tag systems with deletion number $v$ and $\mu$ symbols. Now, let $T$ be a $v$-tag system in the class TS($\mu, v$) and appendants $w_{0}, w_{1},\ldots,w_{\mu - 1}$. Then:\\
a. $l_{A}$ denotes the length of word $A$.\\
b. $l_{\mathbf{max}}$ denotes the length of the lengthiest appendant $w_{i}$, $l_{\mathbf{min}}$ the length of the shortest appendant $w_{j}$, $0 \leq i, j < \mu$.\\
c. $T$ is said to have \textit{unbounded growth} on word $A_{0}$ iff. for each natural number $n$ there exists an $i$ such that for each $j > i$, any word $A_{j}$ produced after $j$ computation steps of $T$ on $A_{0}$ has length greater than $n$.\\
d. A word $A = a_{1}a_{2}\ldots a_{l_{A}}$ is said to be \textit{a periodic word} with period $p$ if there is a $p$ such that $T$ will reproduce $A$ after $p$ computation steps of $T$ starting from $A$.\\
e. The set of words $[P]$ is called \textit{a set of periodic words} with period $p$ when for any $P_{i}, P_{j} \in [P]$, $1 \leq i,j \leq p$, $T$ produces $P_{j}$ from $P_{i}$ after at most $p$ computation steps. \\
f. The \emph{periodic structure}, $\mathfrak{S}$, of a periodic word $S = a_{1, S}a_{2, S}\ldots a_{l_{S}, S}$ is the word\linebreak $a_{1, S}a_{v+1, S}a_{2v + 1}\ldots a_{l_{S} - (l_{S} - 1 \bmod v), S}$. 
\section{Theoretical Methods and Results.}\label{TheorResults}
\subsection{Decidability Criteria for Tag Systems}\label{criteria}
The results by Cocke and Minsky (Sec. \ref{Intro}) were generalized by Maslov. He proved that for any $v > 1$ there is at least one tag system with an unsolvable decision problem \cite{masl64b} and furthermore proved that any tag system with $v = 1$ has a recursively solvable halting and reachability problem. This was proven independently by Wang \cite{wang63}, and about 40 years earlier by Post. It thus follows that the deletion number $v$ is one decidability criterion \cite{marg00} for tag systems with $v = 2$ as the frontier value.\\
Another such criterion is the length of the appendants. Wang proved that any tag system with
$l_{\mathbf{min}} \geq v$ or $l_{\mathbf{max}} \leq v$ has a solvable halting and reachability problem
\cite{wang63}. He also proved that there is a universal tag system with $v = 2,  l_{\mathbf{max}} = 3,  l_{\mathbf{min}} = 1$. This result was proven independently by Maslov \cite{masl64b}. Minsky and Cocke also constructed a universal tag system with the same parameters \cite{CoMinsk63}. This criterion was also studied by Pager \cite{pag70}. It follows from these results that $l_{\mathbf{max}} - v$
resp. $v - l_{\mathbf{min}}$ are decidability criteria for tag systems with 1 as the frontier value. \\
A third decidability criterion is the number of symbols $\mu$. Post proved that the classes TS$(1, v)$ and TS(2, 2) have a solvable halting and reachability problem. Regretfully, Post never published these results. He does mention that the proof for the class TS(1, $v$) is trivial, while the proof for the class TS(2,2) involved ``\textit{considerable labor}''. A proof for the class TS(2,2) has recently been established. The proof is quite involved due to the large number of cases studied. An outline of the proof as well as a description of the main method of proof, called the table method, can be found in \cite{DeMol07}. Although it follows from these results that $\mu$ is a decidability criterion, its frontier value is still unknown.\\
Until recently the number of symbols $\mu$ was never really studied, with Post as an exception. As a consequence, although one has constructed the smallest possible universal tag systems with respect to $v,  l_{\mathbf{max}}$ and  $l_{\mathbf{min}}$, the value of $\mu$ for these universal tag systems is still relatively large. 
It follows from the results of  \cite{CoMinsk63,minsk62}  that it is possible to reduce any 2-symbolic Turing machine with
$m$ states to a tag system with $v = 2$, $\mu = 32m$. Using the
universal Turing machine constructed by Neary and Woods in the class TM(15,
2) which simulates a variant  of tag systems called bi-tag systems \cite{near06} or the machine constructed by Baiocchi, which simulates 2-tag systems, in the class TM(19,2) \cite{bai01}, where TM($m,n$) denotes the
class of Turing machines with $m$ states and $n$ symbols, it is
possible to construct universal tag systems in the classes TS(480,
2) resp. TS(608,2). If one would be able to construct a universal tag system with a \textit{significantly} smaller number of symbols $\mu$ it might be the case that $v,  l_{\mathbf{max}}$ and  $l_{\mathbf{min}}$ would become significantly larger. I.e., it might be the case that there is a trade-off between $\mu, v,  l_{\mathbf{max}}$ and  $l_{\mathbf{min}}$ comparable to the trade-off between number of states and number of symbols in Turing machines.

\subsection{Reduction of the Collatz problem to TS(3,2).}
Let $C: \mathbb{N} \rightarrow \mathbb{N}$ be defined by:
\begin{displaymath} C(n) = \left\{ \begin{array}{ll} \frac{n}{2} & \textrm{if n is even}\\
3n + 1 & \textrm{if n is odd}\end{array} \right. \end{displaymath}
The Collatz problem is the problem to determine for any $n \in
\mathbb{N}$, whether $C(n)$ will end in a loop $C(4) = 2, C(2) =
1, C(1) = 4$, after a finite number of iterations. For now, the conjecture is that any number $n$ will ultimately end in this loop. This has been checked for all starting values up to $10 \times 2^{58}$.\\
In \cite{DeMol08} it was proven that the Collatz problem can be reduced to a tag system with $v = 2$, $\mu = 3$. The production rules are: $a_{0} \rightarrow a_{1}a_{2}, a_{1} \rightarrow a_{0},
a_{2} \rightarrow a_{0}a_{0}a_{0}$. Note that  $l_{\mathbf{max}} - v = v - l_{\mathbf{min}} = 1$.\\
The Collatz problem had previously already been reduced to several small Turing machines by Baiocchi, (mentioned in \cite{marg00}), Margenstern \cite{marg00}
and Michel \cite{mich93}. However, the description of these machines is longer than that of the Collatz tag system.\\
These results are very significant because the $3n+1$-problem is known to be a very intricate problem of number theory.  Given the complexities involved with the Collatz problem, it might thus be very hard to prove the class TS(3, 2) recursively solvable.  

\subsection{Fast universal tag systems.}
As was explained in Sec. \ref{Intro}, many known small universal systems, including Turing machines, are simulators of 2-tag systems. However, since Cocke and Minsky's  2-tag simulation of Turing machines is exponentially slow, every one of these universal systems suffered from this same defect and it was thus unclear whether or not there is a trade-off between program size and space/time complexity. However, Neary and Woods proved quite recently that 2-tag systems are efficient simulators of Turing machines. They proved that cyclic tag systems, a variant of tag systems, simulate Turing machines in polynomial time and that 2-tag systems are efficient simulators of cyclic tag systems \cite{NearWoods06b,NearWoods06c}. These results not only prove that there are fast universal tag systems, but, more importantly, they imply that many known small universal systems are polynomial time simulators. 
\section{Heuristic Methods and Results.}\label{HeuristicResults}
\begin{quote}
''Post found this (00, 1101) problem ``intractable'', and so did I even with the help of a computer. Of course, unless one has a theory, one cannot expect much help from a computer (unless \textit{it} has a theory) except for clerical aid in studying examples; but if the reader tries to study the behaviour of 100100100100100100100 without such aid, he will be sorry.''\\
\begin{flushright}Marvin Minsky, 1967.\end{flushright}
\end{quote}
Upon reading Post's description of his research on tag systems, it becomes clear that he tested several different tag systems, trying out several initial words for each of these tag systems, in order to come to a better understanding of these systems, initially, with the hope of proving them recursively solvable. Nowadays this kind of approach can be seriously extended through the use of computer experiments. Computer experiments can indeed be useful tools to study small computational systems. The research of Wolfram on cellular automata is one more well-known example of this kind of approach \cite{wolf02}. Another example is the use of the computer to find winners of the Busy beaver game (see e.g. \cite{rad62}). Computer experiments can help to build up an intuition of a given class of computational systems, they can suggest new approaches and conjectures, and, in some cases, might even result in a proof of a certain conjecture.  One should always be extremely careful both in the set-up of and the process of drawing conclusions on the basis of computer experiments. Also the formulation of a conjecture should always be done with extreme care. 
\subsection{Post's tag system in TS(2,3).}\label{PostsTag}
Probably the most well-researched tag system is the one example provided by Post with $v = 3, \mu = 2, 0 \rightarrow 00, 1 \rightarrow 1101$. Up to today, it is still not known whether this particular example has a decidable reachability problem, despite its apparent simplicity. 
Post mentions that the numerous initial words he tested always led to a halt or to periodicity and that he had a kind of statistical method to predict that an initial word would become periodic. After Post some other researchers, including Minsky \cite{minsk67}, tested this tag system on their computer, resulting in the same conclusion: every initial word tested resulted in a halt or periodicity. It is mentioned by Asveld \cite{Asveld1996} that some people claimed to have found a counterexample, i.e., the word $A_{0} = (100)^{110}$ would lead to unbounded growth. Asveld disproved this. After 43 913 328 040 672 computation steps, the tag system halts on this initial word.\\ 
This kind of ultimate behaviour does not mean that this tag system is a very easy one to understand. On the contrary, if one observes the computational process of this tag system the evolution of the lengths of the words over time is highly erratic. Furthermore, the fact that the tag system behaves erratically but, in the end, always seems to result in a halt or periodicity together with the fact that it is not unrealistic to assume that the Collatz problem can be reduced to a very small tag system, suggests that this tag system might be closely related to the Collatz problem. This has been suggested before by Ansveld and Hayes \cite{Asveld1996,hay86}. Proving this one tag system either solvable or recursively unsolvable could thus have an impact on research on the Collatz problem. An interesting approach for studying this tag system would be to deepen its possible connection with the Collatz problem.\\
Post's tag system was also studied by Watanabe, who is known for his work on constructing small universal Turing machines in the 60s (see e.g. \cite{wat61}). He did not use computer experiments but made a detailed theoretical analysis of the periodic behaviour of the tag system``\textit{as a preliminary of obtaining a simple universal process}'' \cite{wat63}.  Let $a = 00, b = 1101$. Watanabe deduced wrongly that there are only four kinds of periodic words in Post's tag system, i.e., $b^2a^3(b^3a^3)^n$ with period 6, $ba$ with period 2, $b^2a^2$ with period 4, or any concatenation of the last two.
\subsection{Five computer experiments on the class TS(2,$v$)}\label{computerexps}
Six computer experiments were performed on 52 different tag systems in the class TS(2,$v$) \cite{DeMol07PhD} of which we will briefly discuss five here. The main purpose of the experiments was to come to a better understanding of the behaviour of very small tag systems, similar to Post's tag system. The class TS(2,$v$) was chosen because, on the one hand, it contains Post's tag system, and, on the other hand, it allows to study very small tag systems with respect to $\mu$.\\
In the experiments 50 out of the 52 tag systems tested were selected from a randomly generated set of tag systems. The maximum value for $v$ was set to 15. Besides Wang's decidability criterion with $l_{\mathbf{max}} - v \geq 1, v - l_{\mathbf{min}} \geq 1$ the two most important selection criteria used are heuristic in nature. The first criterion concerns the relative proportion between the total number of occurrences 
$\#a_{0}, \#a_{1},\ldots,\#a_{\mu - 1}$ of each of the symbols
$a_{0}, a_{1}, \ldots, a_{\mu - 1}$ in the appendants of a
given tag system $T$. For each symbol $a_{i}$, we can measure the
effect of reading $a_{i}$ on the length of a word $A$ produced by $T$, i.e., it can lead to a decrease, an increase or have no effect on $l_{A}$. This
effect is computed by
$l_{w_{a_{i}}} - v$. If we then sum up the products $\#a_{i} \cdot
(l_{w_{a_{i}}} - v)$ for each of the symbols $a_{i}$, and the result is a
negative resp. a positive number, one might expect that, on the average, $T$ will
halt or become periodic resp. show unbounded growth. Although this is not true in general, the criterion allows for a quick selection of tag systems for which it is not immediately obvious that they have a solvable reachability problem. It should be noted that this criterion is based on the fact that Post's tag system also has this property.\footnote{This was first pointed out by Minsky \cite{minsk67}.}\\ 
The tag systems thus selected were each run with 20 different and randomly selected initial words of length 300. If the tag system did not lead to a halt, periodicity or was not recognized as a possible case of unbounded growth after 10.000.000 computation steps for any of these initial words, it was selected. Since it is very difficult to track unbounded growth, we simply placed a bound on the lengths of the words produced. If the tag system produced a word $W$ with $L_{W} > 15000$ it was excluded.\footnote{Note that this does not necessary mean that the tag system is really a case of unbounded growth. The reason for choosing such a limit is that for those tag systems $T \in TS(2,2)$ that have been proven to show unbounded growth, the length of the words grows very fast. It thus seemed reasonable to assume that if one has a tag system that can be easily proven to show unbounded growth, then the length of the words produced by this tag system will grow very fast. If this is not the case one rather expects that as long as the tag system does not halt or become periodic the average length of the words will increase very slowly.} From the 50 tag systems thus generated, the smallest resp. the largest $v$ was 3 resp. 13. The smallest resp. largest value for $l_{\mathbf{max}} - v$ and  $v - l_{\mathbf{min}}$ was 1 and 4.\\
The other two tag systems tested are, on the one hand, Post's tag system, and, on the other hand, a tag system that we ourselves had already constructed and tested to some extend.
\subsubsection{Experiment 1: Distribution of the three classes of behaviour}
\begin{figure}[h]
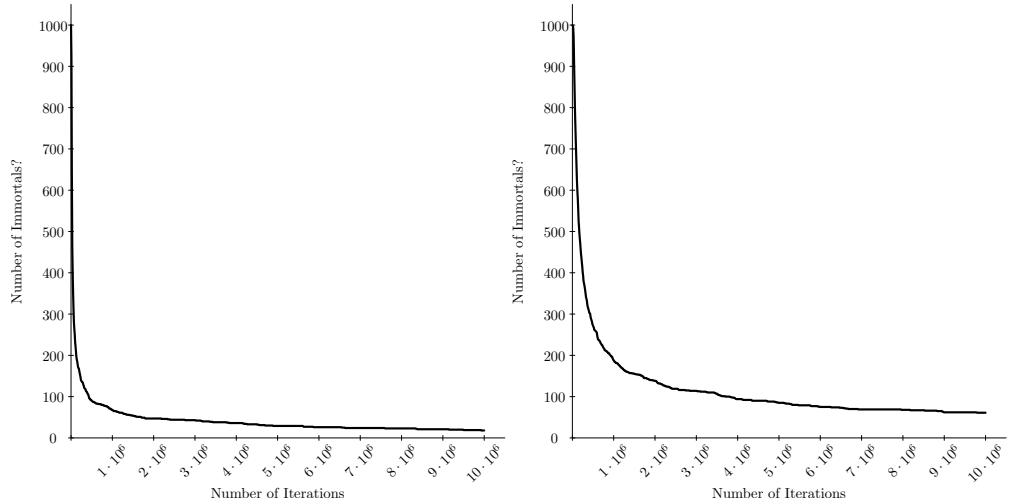

    \label{plots}
    \begin{center}

        \includegraphics[height = 0.3\textheight]{Grapht1-new.pdf}
        \includegraphics[height = 0.3\textheight]{Grapht40-new.pdf}
     \end{center}
 \caption{The plot on the left is the plot for Post's tag system, the plot on the right is the plot of a tag system $v = 5, 0 \rightarrow 1011, 1 \rightarrow 010100$.}
\end{figure}

In the first experiment, each of the 52 different tag systems was run with 1998 randomly selected initial words. The program kept track of the number of initial words leading to a halt, periodicity, the production of a word $W$ with $l_{W} > 15000$ and the number of initial words that did not lead to either of these three classes of behaviour after 10.000.000 computation steps.  The results showed, for each of the tag systems, that the number of initial words that produced words $W$ with $l_{W} > 15000$ was very low, varying between 0 and 43. 
If an initial word did lead to a halt, periodicity or the production of a word $W$, $l_{W} > 15000$, the program also kept track of the number of computation steps it took the tag system before any of these three cases occurred. On the basis of this count, a plot was made for each of the tag systems showing the number of computation steps against the number of initial words that have not led to a halt, periodicity or the production of a word $W$, with $l_{W} > 15000$.\footnote{Note that since the number of initial words that  produced a word $W, l_{W} > 15000$ was for each of the tag systems very small, they did not have a major impact on the form of the plots.}\\
We do not have the space to give all the results here. However, it should be noted that there was a clear variation between the different tag systems concerning the distribution of the number of initial words that halt, become periodic, produced a word $W, l_{W} > 15000$ or those that could not be placed in any of these three classes of behavior. Most significant were the results from the plots. Fig. \ref{plots} shows two such plots.\\

As is clear from these plots, the number of initial words that do not lead to a halt or periodicity, first decreases exponentially fast but then decreases exponentially slow.  Most of the plots for the other 50 tag systems are quite similar to these plots.
Now, if one could generalize these plots, i.e., if one would always get this kind of switch between exponentially fast decrease and exponentially slow decrease in the number of non-halting and non-periodic words, whatever the length and the number of initial words tested, these plots indicate that the 52 tag systems tested behave quite unpredictably. Indeed, this would mean that given any $n$, it is always possible to find an initial word that will not have halted or become periodic after $n$ computation steps. Of course, this is always the case for tag systems that can be proven to show unbounded growth, i.e., tag systems with a decidable reachability problem. However, the results from this and experiments 3--5 make it seem quite improbable that  these tag systems grow in a ``predictable'' manner. Rather the evolution of the lengths of the words over time is highly erratic. More experiments would be needed to support this.    

\subsubsection{Experiment 2: Periodicity in tag systems}\label{periods}
In the second experiment the periodic behaviour of each of the tag systems was studied. Research on periodic behaviour of a certain class of computational systems can be very fruitful. For example, Cook used periodic words to prove that cellular automaton rule 110 is weakly universal \cite{mcook04}.

A detailed analysis was performed on the periods produced by the initial words from experiment 1 that led to periodicity, making use of both the computer as well as the more common pencil-and-paper method.  The most important result from this experiment is that we found four basic types of periods, summarized in Table \ref{types}.

\begin{table}
\begin{center}
\begin{tabular}{|l|c|c|c|c|}
\hline
 & Type 1 & Type 2 & Type 3 & Type 4\\
 \hline
 $\forall S \in [P]:$ min$(p, l_{\mathfrak{S}})$ & $p$ or $p =  l_{\mathfrak{S}}$ & $p$ & $l_{\mathfrak{S}}$ & $l_{\mathfrak{S}}$\\
 \hline
 $\exists S \in [P]: n \stackrel{?}= \frac{\mathrm{max}(p,
 l_{\mathfrak{S}})}{\mathrm{min}(p,
 l_{\mathfrak{S}})}$ & $\checkmark$ & $\diagdown$ & $\checkmark$ &
 $\diagdown$\\
 \hline 
 \end{tabular}
 \end{center}
\caption{Summary of the different types of periods.}\label{types}
 \end{table}
Min$(p, l_{\mathfrak{S}})$ resp. max$(p, l_{\mathfrak{S}})$ gives the minimum resp. maximum of $p$ and $l_{\mathfrak{S}}$ for a periodic word $S \in [P]$. Row 2 gives min$(p, l_{\mathfrak{S}})$ for each of the four types. For a set of periodic words $[P]$ of type 1 or 2, the period $p$ is always smaller than or equal to $l_\mathfrak{S}$ for each $S \in [P]$, for words $[P]$ of type 3 or 4, we have the opposite, i.e.,  $l_\mathfrak{S}$ is always \textit{strictly} smaller than $p$. In row 3, if for a given set of periodic words $[P]$ of type $x$, there is a word $S \in [P]$ for which max$(p, l_{\mathfrak{S}})$ is divisible by min$(p, l_{\mathfrak{S}})$, then $\checkmark$ is used. If this is not the case  $\diagdown$ is used. We will not provide examples here of each of the four types.\footnote{An example of a period of type 1 was given in Sec. \ref{definitions}.} In \cite{DeMol07PhD} types 1 and 3 were called regular types, types 2 and 4 were identified as irregular types. The fundamental difference between regular and irregular types has several consequences. Given two sets of periodic words $[P_{1}]$ and $[P_{2}]$ with period $p_{1}$ resp. $p_{2}$. If the words in $[P_{1}]$ and $[P_{2}]$ are either of type 1 or 3, then there is at least one $W_{i, 1} \in [P_{1}]$ and one $W_{j, 2} \in [P_{2}]$, $1 \leq i \leq p_{1}, 1 \leq j \leq p_{2}$ such that any concatenation of $W_{i, 1}$ and $W_{j, 2}$ is again a periodic word. It is exactly this property that allows to generate certain number sequences with tag systems that are able to produce at least two different sets of periodic words of type 1 or 3. For example, it is possible to produce, for arbitrary $n \in \mathbb{N}$, any period $p = 2n$ with Post's tag system.\footnote{This was first noticed by Shearer \cite{shear96}.}  This kind of construction is impossible with periods of type 2 and 4.

For Post's tag system, only words of type 1 were produced during experiment 1. However, during several preliminary tests on Post's tag system, we also found two different periodic words of type 4. Watanabe never considered the possibility of periodic words of type 4. This is one of the reasons why his conclusions on the periodic behaviour of Post's tag system are incorrect (see Sec \ref{PostsTag}).

Of course, for now, we are not sure whether these are the only possible types of periods or whether the four types found contain important new subtypes. This is one of the typical drawbacks of experimental research. More research in this direction could be important, especially given the fact that these periodic types provide a possible connection between tag systems and number theory. It would e.g. be very interesting to know what kind of different number series could be produced by the periodic words of a given tag system. Furthermore, periods of type 1 and 3 seem very promising with regard to finding small universal tag systems with the help of a computer (Sec. \ref{Conclusion}).

It should also be mentioned here that it can be proven that tag systems $T \in $ TS(2,2) are incapable of producing words of type 2 and 4, i.e., irregular types. This is a fundamental difference between this class, known to have a solvable reachability problem, and the classes TS(2, 3) and TS(3, 2).  

\subsubsection{Experiment 3 -- 5: Summary of the results.} 
In the three remaining experiments, we tried to measure how unpredictable these 52 tag systems actually are. The results of experiment 3 showed that each of the tag systems has a very high sensitivity to the initial words. I.e., one small change in the initial word led to large variations in the long-term behavior of the tag systems. This is a sign of chaotic behavior. Experiment 4 checked whether the distribution of the 0s and 1s read in the words produced by these tag systems is random or not. For this experiment we used Marsaglia's DIEHARD, a battery of tests for randomness (stat.fsu.edu/pub/diehard/). Although none of the tag systems passed all tests, there were only two that passed none of the tests, one of them being Post's tag system. This rather came as a surprise since Post's tag system is known to behave very erratically. It indicates that there might be an important difference between this tag system and 50 of the other tag systems. We have not been able yet to study this in more detail. In the last experiment we performed a Markov analysis on the words produced by these tag systems in order to compute their information-theoretical entropy as defined by Claude Shannon \cite{shan48}. For most of the tag systems, this entropy was very high, some were even very close to the maximum value 1.0.\\

\noindent As is clear from these 5 experiments, the class of tag systems with $\mu = 2, 2 < v < 15, 1 \leq 
l_{\mathbf{max}} - v, v - l_{\mathbf{min}} \leq 4$ contains tag systems that behave quite unpredictably. This feature  serves as an indication that there are very small tag systems for which it might be very hard if not impossible to prove them recursively solvable. 
\section{Discussion. In search of small universal tag systems.}\label{Conclusion}
As was explained in Sec. \ref{criteria}, there are three known decidability criteria for tag systems, i.e., $v, l_{\mathbf{max}} - v, v - l_{\mathbf{min}} $ and $\mu$. The frontier value for the first two is known, the frontier value for $\mu$ is unknown.  As a consequence, there is a significant gap between the size of the known universal tag systems and the tag systems known to have a decidable reachability problem. Despite the relatively large size of $\mu$ for the known universal tag systems, there are some clear indications that it might be possible to decrease $\mu$. The fact that the Collatz problem can be reduced to a tag system in TS(3, 2) is one such indication. The experimental results from Sec.\ref{HeuristicResults}, serve as another such indication as they show that very small tag systems show very complicated behavior.\\
There are two main approaches to tackle the problem of finding smaller universal tag systems. The first, more theoretical, approach, is to search for a simulation of a known universal class of computational systems by a tag system, trying to decrease $\mu$ and keeping $v$ and $l_{\mathbf{max}} - l_{\mathbf{min}}$ relatively small. This is the standard approach in the context of finding small universal systems in other models of computability like e.g. Turing machines. It seems quite probable that this kind of approach will lead to smaller universal tag systems, since, until now, nobody has really searched for small universal tag systems relative to $\mu$.\\
A second, more heuristic, approach is to start from a detailed computer analysis of a specific class of tag systems for which there are indications that it might contain a tag system with a recursively unsolvable problem. This can be a very fruitful approach, as is e.g. clear from Cook's proof that cellular automaton rule 110 is able to simulate any cyclic tag system, which is based on a detailed analysis of the behaviour of rule 110 \cite{mcook04,wolf02}. In fact, a detailed computer analysis of one of the periodic words of type 1 in Post's tag system, using the table method mentioned in Sec. \ref{criteria}, has shown that it might be possible to simulate the Cocke-Minsky encoding by making use of periods of type 1 (see Sec. \ref{periods}) \cite{DeMol07PhD}. I.e., the computer analysis shows that one can do certain manipulations on these periodic words which correspond to several of the kind of operations needed in the Cocke-Minksy scheme, like e.g. halving or doubling the length of a subword. However, more research is needed here. One of the major problems that still has to be solved is that one needs to find a way to synchronize every one of the individual operations on the periodic words.\\
As is clear from the results discussed in Sec. \ref{TheorResults} and \ref{HeuristicResults} both approaches complement each other. For example, the simulation of the Collatz problem in a very small tag system together with the more experimental results discussed in Sec. \ref{computerexps}, have provided us with more information on very small classes of tag systems. Especially in research on very small tag systems like e.g. Post's tag system a combined approach seems the most promising.\\
In recent years there has been some research on non-standard or more general definitions of universality in the context of cellular automata and Turing machines (see e.g. \cite{mcook04,WoodsNearyMCU07}). By using more general definitions of universality it is possible to lower the boundaries of undecidability for these computational models. We expect that 
if it would be possible to prove very small classes of tag systems universal, one would need a similar more general notion of universality for tag systems. One possibility is to consider tag systems that cannot halt but always either show unbounded growth or become periodic. \\ 

\newcommand{\noopsort}[1]{}
\providecommand{\bysame}{\leavevmode\hbox to3em{\hrulefill}\thinspace}
\providecommand{\MR}{\relax\ifhmode\unskip\space\fi MR }
\providecommand{\MRhref}[2]{%
  \href{http://www.ams.org/mathscinet-getitem?mr=#1}{#2}
}
\providecommand{\href}[2]{#2}
\bibliographystyle{eptcs}


\begin{thebibliography}{}
\providecommand{\bibitemstart}[1]{\bibitem{#1}}
\providecommand{\bibitemend}{}
\providecommand{\bibliographystart}{}
\providecommand{\bibliographyend}{}
\providecommand{\url}[1]{\texttt{#1}}
\providecommand{\urlprefix}{Available at }
\providecommand{\bibinfo}[2]{#2}
\bibliographystart

\bibliographyend
\end{thebibliography}


\begin{thebibliography}{1}

\bibitem{Asveld1996}
Peter~R.J. Asveld, \emph{An inefficient representation of the empty word},
  Memorandum Informatica 96-14, Department of Computer Science, University of
  Twente, Enschede, 1996.

\bibitem{bai01}
Claudio Baiocchi, \emph{Three small universal Turing machines}, Proc. 3rd
  International Conference on Machines, Computations, Universality (Berlin)
  (Yu.~Rogozhin M.~Margenstern, ed.), Lecture Notes in Computer Science, vol.
  2055, 2001, pp.~1--10.

\bibitem{CoMinsk63}
John Cocke and Marvin Minsky, \emph{Universality of tag systems with p = 2},
  1963, Artificial Intelligence Project -- RLE and MIT Computation Center, memo
  52.

\bibitem{CockeMinsky64}
\bysame, \emph{Universality of tag systems with p = 2}, Journal of the ACM
  \textbf{11} (1964), no.~1, 15--20.

\bibitem{mcook04}
Matthew Cook, \emph{Universality in elementary cellular automata}, Complex
  Systems \textbf{15} (2004), no.~1, 1--40.

\bibitem{hay86}
Brian Hayes, \emph{Theory and practice: Tag-you're it}, Computer Language
  (1986), 21--28.

\bibitem{marg00}
Maurice Margenstern, \emph{Frontier between decidability and undecidability: A
  survey}, Theoretical Computer Science \textbf{231} (2000), no.~2, 217--251.

\bibitem{masl64b}
Sergei.~J. Maslov, \emph{On \uppercase{E}. \uppercase{L}. \uppercase{P}ost's
  `\uppercase{T}ag' problem. (russian)}, Trudy Matematicheskogo Instituta imeni
  V.A. Steklova (1964b), no.~72, 5--56, English translation in: American
  Mathematical Society Translations Series 2, 97, 1--14, 1971.

\bibitem{mich93}
Pascal Michel, \emph{Busy beaver competition and Collatz-like problems},
  Archive for Mathematical Logic \textbf{32} (1993), no.~5, 351--367.

\bibitem{minsk61a}
Marvin Minsky, \emph{Recursive unsolvability of \uppercase{P}ost's problem of
  tag and other topics in the theory of \uppercase{T}uring machines}, Annals of
  Mathematics \textbf{74} (1961), 437--455.

\bibitem{minsk62}
\bysame, \emph{Size and structure of universal Turing machines using tag
  systems: a 4-symbol 7-state machine}, Proceedings Symposia Pure Mathematics,
  American Mathematical Society \textbf{5} (1962), 229--238.

\bibitem{minsk67}
\bysame, \emph{Computation. Finite and infinite machines}, Series in Automatic
  Computation, Prentice Hall, Englewood Cliffs, New Jersey, 1967.

\bibitem{DeMol07}
Liesbeth~De Mol, \emph{Study of limits of solvability in tag systems},
  Machines, Computations, and Universality. Fifth International Conference, MCU
  2007 Orl{\'e}ans (Berlin) (J.~Durand-Lose and M.~Margenstern, eds.), Lecture
  Notes in Computer Science, vol. 4664, Springer, 2007, pp.~170--181.

\bibitem{DeMol07PhD}
\bysame, \emph{Tracing unsolvability: A historical, mathematical and
  philosophical analysis with a special focus on tag systems}, Ph.D. thesis,
  University of Ghent, 2007.

\bibitem{DeMol08}
\bysame, \emph{Tag systems and Collatz-like functions}, Theoretical Computer
  Science \textbf{390} (2008), no.~1, 92--101.

\bibitem{near06}
 Turlough Neary and Damien Woods, \emph{Four small universal Turing machines},
Fundamenta Informaticae, 91(1):123--144, 2009.

%

\bibitem{NearWoods06b}
\bysame, \emph{P-completeness of cellular automaton rule 110}, International
  Colloquium on Automata Languages and Programming (ICALP), Lecture Notes in
  Computer Science, vol. 4051, {\noopsort{a}}2006, pp.~132--143.

\bibitem{pag70}
David Pager, \emph{The categorization of tag systems in terms of decidability},
  Journal of the London Mathematical Society \textbf{2} (1970), no.~2,
  473--480.

\bibitem{post43}
Emil~Leon Post, \emph{Formal reductions of the general combinatorial decision
  problem}, American Journal of Mathematics \textbf{65} (1943), no.~2,
  197--215.

\bibitem{post65}
\bysame, \emph{Absolutely unsolvable problems and relatively undecidable
  propositions - account of an anticipation}, The Undecidable.
  \uppercase{b}asic papers on undecidable propositions, unsolvable problems and
  computable functions (Martin Davis, ed.), Raven Press, 1965, pp.~340--433.

\bibitem{rad62}
Tibor R{\'a}do, \emph{On non-computable functions}, The Bell System Technical
  Journal \textbf{41} (1962), no.~3, 877--884.

\bibitem{rog82}
Yurii Rogozhin, \emph{Seven universal Turing machines (in russian)}, Mat.
  Issledovaniya \textbf{69} (1982), 76--90.

\bibitem{shan48}
Claude~E. Shannon, \emph{A mathematical theory of communication}, Bell System
  Technical Journal \textbf{27} (1948), 379--423 en 623--656.

\bibitem{shear96}
James~B. Shearer, \emph{Periods of strings (letter to the editor)}, American
  Scientist \textbf{86} (1996), 207.

\bibitem{wang63}
Hao Wang, \emph{Tag systems and lag systems}, Mathematische Annalen
  \textbf{152} (1963a), 65--74.

\bibitem{wat61}
Shigeru Watanabe, \emph{5-symbol 8-state and 5-symbol 6-state universal Turing
  machines}, Journal of the ACM \textbf{8} (1961), no.~4, 476--483.

\bibitem{wat63}
\bysame, \emph{Periodicity of Post's normal process of tag}, Mathematical
  Theory of Automata (Brooklyn, NY) (Jeremy Fox, ed.), Microwave Research
  Institute Symposia Series, vol. XII, Polytechnic Press, 1963, pp.~83--99.

\bibitem{wolf02}
Stephen Wolfram, \emph{A new kind of science}, Wolfram Inc., Champaign, 2002.

\bibitem{WoodsNearyMCU07}
Damien Woods and Turlough Neary, \emph{Small semi-weakly universal Turing
  machines}, Machines, Computations, and Universality. Fifth International
  Conference, MCU 2007 Orl{\'e}ans (J.~Durand-Lose and M.~Margenstern, eds.),
  LNCS, vol. 4664, Springer, 2007, pp.~303--315.

\bibitem{NearWoods06c}
\bysame, \emph{On the time complexity of 2-tag systems and small universal
  Turing machines}, Proceedings of the 47th Annual IEEE Symposium on
  Foundations of Computer Science, {\noopsort{b}}2006, pp.~439--448.

\end{thebibliography}
\end{document}